\documentclass[prb,aps,twocolumn,draft,tightenlines,%
showpacs,floatfix,superscriptaddress]{revtex4}
\usepackage{amssymb}
\usepackage{psfig}
\usepackage{amsmath}
\usepackage{colordvi}
\begin{document}

\title{Spin relaxation in $n$-type GaAs quantum wells from a
fully microscopic approach}
\author{J. Zhou}
\affiliation{Hefei National Laboratory for Physical Sciences at
  Microscale, University of Science and Technology of China,
Hefei, Anhui, 230026, China}
\affiliation{Department of Physics, University of Science and
Technology of China, Hefei, Anhui, 230026, China}
\altaffiliation{Mailing Address.}
\author{J. L. Cheng}
\affiliation{Department of Physics, University of Science and
Technology of China, Hefei, Anhui, 230026, China}
\altaffiliation{Mailing Address.}
\author{M. W. Wu}
\thanks{Author to whom all correspondence should be addressed}
\email{mwwu@ustc.edu.cn.}
\affiliation{Hefei National Laboratory for Physical Sciences at
  Microscale, University of Science and Technology of China,
Hefei, Anhui, 230026, China}
\affiliation{Department of Physics, University of Science and
Technology of China, Hefei, Anhui, 230026, China}
\altaffiliation{Mailing Address.}
\date{\today}

\begin{abstract}

We perform a full microscopic investigation
on the spin relaxation in $n$-type (001) GaAs quantum wells
with Al$_{0.4}$Ga$_{0.6}$As barrier
due to the D'yakonov-perel' mechanism from nearly 20\ K to
the room temperature by constructing and numerically
solving the kinetic spin Bloch equations.
We consider all the relevant scattering such as the electron--acoustic-phonon,
the electron--longitudinal-optical-phonon,
the electron--nonmagnetic-impurity and the
electron-electron Coulomb scattering to the spin relaxation.
The spin relaxation times calculated from our
theory with a fitting spin splitting parameter
are in good agreement with the experimental data
by Ohno {\em et al.} [Physica E {\bf 6}, 817 (2000)] over the
whole temperature regime (from 20\ K to 300\ K).
The value of the fitted spin splitting parameter
agrees with many experiments and theoretical calculations.
We further show the temperature dependence of the spin relaxation time
under various conditions such as  electron density, impurity density
and well width.  We predict a peak solely due to the Coulomb scattering
in the spin relaxation time at low temperature ($<50$\ K)
in samples with low electron density ({\em e.g.}, density less than
$1 \times 10^{11}$\ cm$^{-2}$)
but high mobility.
This peak disappears in samples with high
electron density ({\em e.g.} $2 \times 10^{11}$\ cm$^{-2}$)
and/or low mobility. The hot-electron spin kinetics at low temperature
is also addressed with many features quite different from the high temperature
case  predicted.
\end{abstract}

\maketitle

\section{INTRODUCTION}

Much attention has been devoted to the electron spin dynamics in semiconductors
for the past three decades.\cite{meier,prinz} Especially, recent
experiments have shown extremely long spin lifetime (up to
hundreds of nanoseconds) in $n$-type bulk Zinc-blende semiconductors
(such as GaAs).\cite{dzhioev,kikka,murdin}
Moreover, a lot more investigations have been performed on
various low dimensional
systems,\cite{damen,heberle,crooker,kikka1,ohno2,ohno,harley2,doh,ohno1,harley1,wagner,lombez,shaff,strand,holl}
and  spin lifetime as long as tens of nanoseconds has been reported
in (110)-oriented GaAs quantum wells (QWs)\cite{ohno1,harley1}
at room temperature and in $p$-type
GaAs:Be/Al$_x$Ga$_{1-x}$As double hetero-structures\cite{wagner} at
low temperature. In these studies, understanding the spin relaxation/dephasing
(R/D) mechanism is one of the most important problems as it is the
prerequisite for the application of the
spintronic devices. It is understood that the D'ayakonov-Perel' (DP)
mechanism is the leading spin R/D mechanism
in $n$-type Zinc-blende semiconductors.\cite{dp} This mechanism
is composed of the contribution from the Dresselhaus term,\cite{dress}
which is due to the lack of inversion symmetry
in the Zinc-blende crystal Brillouin zone (sometimes referred to
as the bulk inversion asymmetry), and that from the Rashba term,\cite{rashba}
which originates from the asymmetric potential within a QW
along the growth direction (sometimes referred to as the structure inversion
asymmetry). Both appear as effective magnetic fields.
For narrow (001) GaAs QW without the additional large bias
voltage, the Dresselhaus term is the leading term:\cite{dp2,car}

\begin{eqnarray}
\label{omegax}
\Omega_x({\bf k})&=&\gamma k_x(k_y^2-\langle k_z^2\rangle), \\
\Omega_y({\bf k})&=&\gamma k_y(\langle k_z^2\rangle-k_x^2),  \\
\label{omegaz}
\Omega_z({\bf k})&=&0\ ,
\end{eqnarray}
in which $\langle k_z^2\rangle$ represents the average of the operator
$-(\partial/\partial z)^2$ over the electronic state of the lowest subband.
Under the finite square well assumption,
 \begin{equation}
\langle k_z^2\rangle=\frac{4A}{a^2}(2\beta+\frac{\xi^2}{\cos^2\xi})\ ,
\end{equation}
where $\xi$ and $\beta$ are the lowest energy solutions of the equations
\begin{eqnarray}
\beta&=&\xi\tan\xi\nonumber\\
\xi^2+\beta^2&=&m^\ast Va^2/2\hbar^2,
\end{eqnarray}
and $A=(\frac{1}{\beta}+\frac{1}{\cos^2 \xi}+\frac{\beta}{\xi^2})^{-1}$
with $V$, $a$ and $m^\ast$ denoting the well depth, well width and
 the effective mass respectively,
In the limiting case, $\lim_{V\to \infty}\langle k_z^2\rangle=(\pi/a)^2$.
$\gamma$ is the spin splitting parameter.\cite{meier} There are a
lot of theoretical
investigations on the spin R/D due to the DP
mechanism lately.\cite{lau,single,kainz} Most of them are within the
framework of single-particle approximation and the Coulomb scattering is
thought to be irrelevant in the spin R/D.

Recently Wu {\em et al.} performed a full microscopic
investigation on the spin R/D and showed that the single-particle
approach is inadequate in accounting for the spin
R/D.\cite{wu1,wu2,wu3,wu4,wu5,wu6,wu7,wu8,wu9,wu10,lue,wu12}
In this approach, the momentum dependence of the effective magnetic field
(the DP term)  and the momentum
dependence of the spin diffusion rate in the direction of
the spacial gradient\cite{wu6} or even the
random spin-orbit interaction\cite{ya}
serve as inhomogeneous broadening.\cite{wu2,wu3} In the presence of
the inhomogeneous broadening, any scattering (even the spin-conserving
scattering), including the Coulomb scattering,\cite{wu2,wu7,wu8,lue}
can cause irreversible dephasing.
Moreover, this approach also includes the counter effect of the
scattering to the inhomogeneous broadening, {\em i.e.},
the suppression of the inhomogeneous broadening by the scattering.
Finally, this approach is valid not only near the equilibrium, but also
far away from the equilibrium,\cite{wu7,wu8} and is applicable to both
the strong  ($|\mathbf{\Omega}|\tau_{p} \ll 1$)
and the weak ($|\mathbf{\Omega}|\tau_{p} \gg 1$) scattering
limits,\cite{lue,wu12} with $\tau_p$ representing
the momentum relaxation time. In the weak scattering limit, the counter
effect of the scattering is less important and adding additional
scattering (including the Coulomb scattering) causes stronger spin
R/D. Whereas in the  strong scattering limit, adding additional scattering
always increases the spin R/D time. The feature is more complicated
when $|\mathbf{\Omega}|\tau_p\sim1$.\cite{lue}
In above studies,\cite{wu1,wu2,wu3,wu4,wu5,wu6,wu7,wu8,wu9,wu10,lue,wu12}
we have been focusing on the high  temperature regime ($T\ge 120$\ K)
where the electron-acoustic (AC) phonon scattering, which is more complicated
in numerical calculation than the electron-longitudinal
optical (LO) phonon scattering, is negligible.
In this paper, we extend the scope of our approach to study  the
spin kinetics at low temperature regime by
including the electron-AC phonon scattering.
 Moreover, we compare
the spin relaxation time (SRT) obtained from our theory with the
experimental data over a wide temperature regime and show the
excellent agreement of our theory with the experiment. We further
show that the Coulomb scattering is important to the spin R/D not
only at high temperatures,\cite{wu7,wu8,lue} but also at low temperatures.
The electron density, impurity density, well width, temperature and
electric field dependences of the SRT are studied in detail.

This paper is organized as follows: In Sec.\ II we set up the model and
give the kinetic spin Bloch equations.
In Sec.\ III we compare our results with the experimental data.
Then, we investigate the
temperature dependence of the spin relaxation under
different conditions such as
electron densities, impurity densities and  well widths in Sec.\ IV.
The effect of Coulomb scattering is also addressed.
The hot-electron effect in spin relaxation is investigated in Sec.\ V.
We summarize in Sec.\ VI.

\section{KINETIC SPIN BLOCH EQUATIONS}

We start our investigation from an $n$-type GaAs (001) QW with
the growth direction along the $z$-axis.
A moderate magnetic field $B$ is applied along the $x$-axis (in
the Voigt configuration). The kinetic spin Bloch equations
can be constructed by using the nonequilibrium Green
function method:\cite{haug}
\begin{equation}
\dot{\rho}_{{\bf k},\sigma \sigma^{\prime}}-e{\bf E}\cdot{\bf \nabla}_{\bf
k}{\rho}_{{\bf k},\sigma \sigma^{\prime}}=\dot{\rho}_{{\bf k},
\sigma \sigma^{\prime}}|_{coh}
+\dot{\rho}_{{\bf k},\sigma \sigma^{\prime}}|_{scatt},
\label{bloch}
\end{equation}
with $\rho_{{\bf k},\sigma \sigma^{\prime}}$ representing the
single particle density matrix elements.
The diagonal and off-diagonal
elements give the electron distribution functions $f_{{\bf k}\sigma}$ and the spin
coherence $\rho_{{\bf k},\sigma-\sigma}$. The second term
in Eq.\ (\ref{bloch}) describes the
energy input from the external electric field ${\bf E}$.
The coherent terms $\dot{\rho}_{k,\sigma \sigma^{\prime}}|_{coh}$ describe the precession
of the electron spin due to the applied magnetic field $B$ and the effective magnetic field
$\mathbf{\Omega}({\bf k})$ [Eqs.\ (\ref{omegax}-\ref{omegaz})] as well as the
effective magnetic field from the Hartree-Fock Coulomb interaction.\cite{wu7}
$\dot{\rho}_{k,\sigma \sigma^{\prime}}|_{scatt}$ in Eq.\ (\ref{bloch})
denote the electron-LO-phonon, the electron-AC-phonon,
the electron-nonmagnetic impurity and the electron-electron Coulomb scattering.
Their expressions are given in detail in Ref.\ \onlinecite{wu8}, except the
additional matrix elements of the electron-AC-phonon scattering.
For the electron-AC-phonon scattering due to the deformation potential,
the matrix elements are given by
$g^2_{{\bf Q},def}=\frac{\hbar\Xi^{2} Q}{2dv_{sl}}|I(iq_z)|^2$,\cite{vogl}
and for the scattering due to the
piezoelectric coupling, the matrix elements read
$g^2_{{\bf Q},pl}=\frac{32\pi^{2}\hbar e^{2} e_{14}^{2}}{\kappa_0^{2}}
\frac{(3q_{x}q_{y}q_{z})^2}{dv_{sl}Q^7}|I(iq_z)|^2$ for the longitudinal
phonon and $g^2_{{\bf Q},pt}=\frac{32\pi^{2}\hbar e^{2} e_{14}^{2}}
{\kappa_0^{2}}\frac{1}{dv_{st}Q^{5}}
(q_{x}^{2}q_{y}^{2}+q_{y}^{2}q_{z}^{2}+q_{z}^{2}q_{x}^{2}
-\frac{(3q_{x}q_{y}q_{z})^2}{Q^2})|I(iq_z)|^2$ for the transverse
phonon.\cite{mahan}
Here ${\bf Q}\equiv({\bf q},q_z)$;
$Q=\sqrt{q_x^2+q_y^2+q_z^2}$; $\Xi=8.5$\ eV is the
deformation potential;
$d=5.31$\ g/cm$^3$ is the mass
density of the crystal; $v_{sl}=5.29\times 10^3$\ m/s ($v_{st}
=2.48\times 10^3$\ m/s) is the velocity of the
longitudinal (transverse) sound wave; $\kappa_0=12.9$ denotes the
static dielectric constant;
and $e_{14}=1.41\times 10^9$\ V/m represents the piezoelectric
constant.\cite{land}
The AC phonon spectra $\omega_{{\bf Q}\lambda}$ are given by
$\omega_{{\bf Q}l}=v_{sl}Q$ for the longitudinal mode and
$\omega_{{\bf Q}t}=v_{st}Q$ for the transverse mode.
The form factor is
\begin{eqnarray}
|I(iq_z)|^2&=&A^2\Big\{\frac{4\beta\cos y-y\sin y}{4\beta^2+y^2}+\frac{1}{\cos^2\xi} \nonumber \\
&&\times[ \frac{\sin y}{y}+\frac{\sin(y+2\xi)}{2y+4\xi}+\frac{\sin(y-2\xi)}{2y-4\xi}]\Big\}^2
\end{eqnarray}
with $y=q_za/2$.  The numerical schemes of the electron-electron Coulomb,
the electron-impurity as
well as the electron-LO phonon scattering have been given in detail in
Ref.\ \onlinecite{wu8}, whereas the numerical scheme for
the electron-AC-phonon scattering is presented in Appendix\ A.
The electron--interface-phonon scattering is negligible due to the thick 
GaAlAs barrier. In addition, as we are going to explore 
the spin R/D over a wide
range of electron densities, in the present paper
we use the screening under the random phase
approximation \cite{RPA}
rather than the one in the
limiting (degenerate or nondegenerate) cases for the screened Coulomb
potential:\cite{haug1}
\begin{equation}
\bar{v}_{q}=\frac{\sum_{q_z}v_{Q}|I(iq_z)|^2}{1-\sum_{q_z}
v_{Q}|I(iq_z)|^2P^{(1)}({\bf q})},
\end{equation}
where $v_{Q}=4\pi e^2/Q^2$ is the bare Coulomb potential and
\begin{equation}
P^{(1)}({\bf q})=\sum_{{\bf k},\sigma}\frac{f_{{\bf k}+{\bf q}\sigma}
-f_{{\bf k}\sigma}}{\epsilon_{\bf k+q}-\epsilon_{\bf k}}\ .
\end{equation}
In this way, we also take into account the hot-electron effect on the
screening.

By numerically solving the kinetic spin Bloch equations with all
these scattering explicitly included, one is able to obtain the
spin dephasing and relaxation times from the
temporal evolutions of the spin coherence $\rho_{{\bf k},\sigma-\sigma}$
and the electron distribution functions $f_{{\bf k},\sigma}$.
The irreversible spin dephasing time can be obtained
by the slope of the envelope of the incoherently summed spin
coherence\cite{wu1}
${\rho}=\sum_{{\bf k}}|{\rho}_{{\bf k},\uparrow\downarrow}(t)|$,
and the SRT can be defined by the slope of the envelope of
the difference between $n_{\uparrow}$ and $n_{\downarrow}$,
with  $n_{\sigma}=\sum_{{\bf k}}f_{{\bf k},\sigma}$.

\begin{figure}[htb]
\psfig{figure=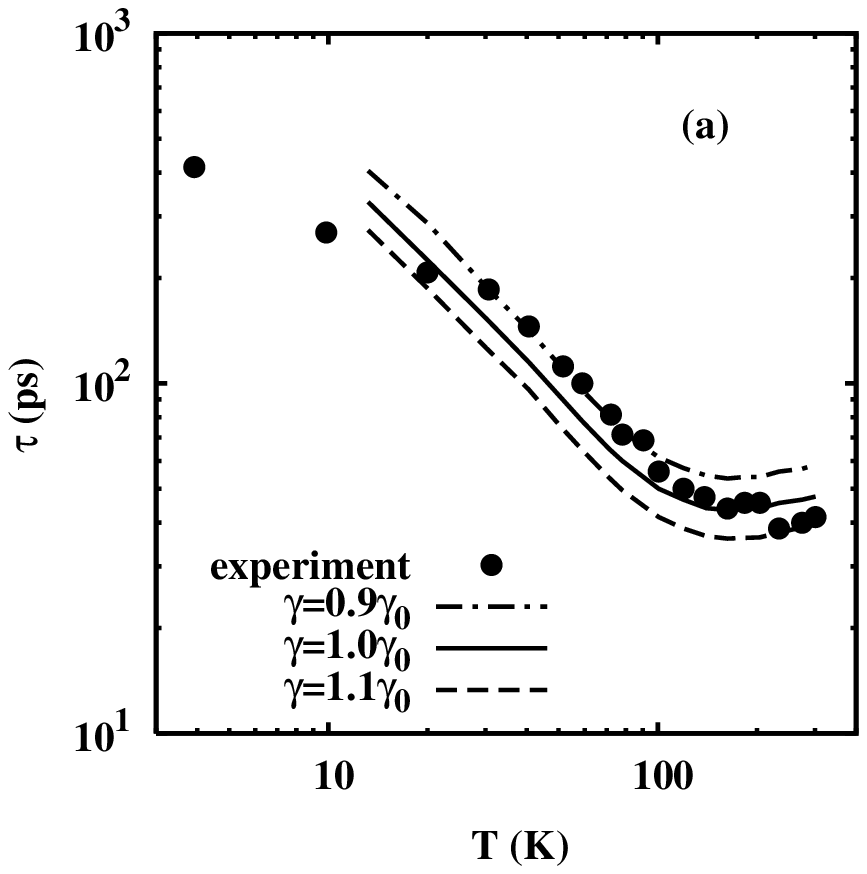,width=8cm,height=5cm,angle=0}
\hspace{-0.4cm}
\psfig{figure=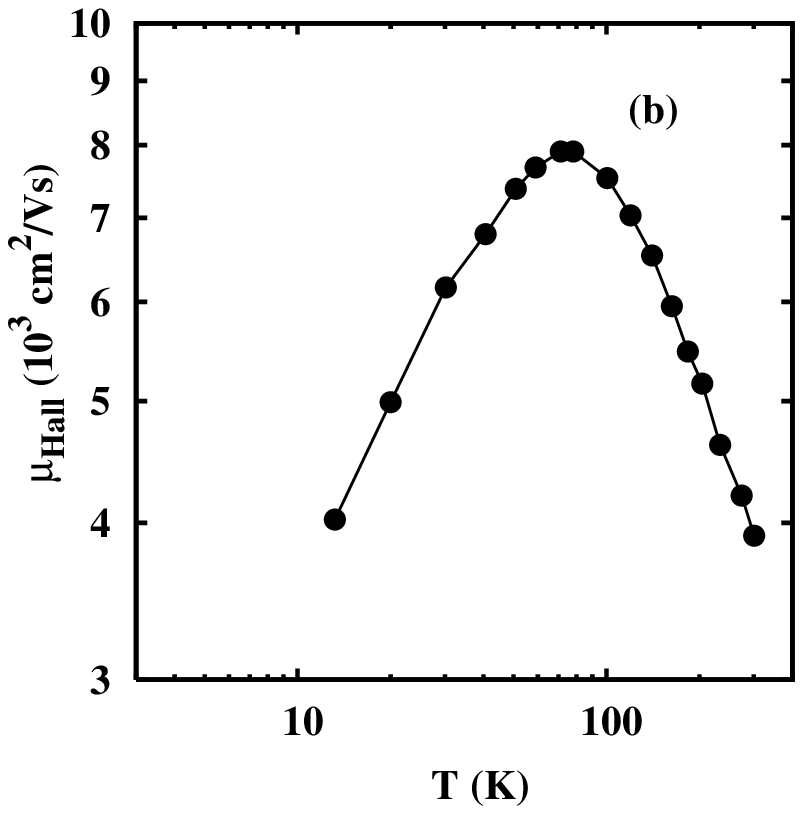,width=8cm,height=5cm,angle=0}
\caption{(a) SRT $\tau$ {\em vs.} temperature $T$ for
GaAs QW with  $a = 7.5$\ nm and electron density $n=4\times
10^{10}$\ cm$^{-2}$ at three different spin-splitting parameters.
Dots: experiment data; Dot-dashed curve: $\gamma = 0.9 \gamma_0$;
Solid curve: $\gamma = \gamma_0$; Dashed curve: $\gamma = 1.1\gamma_0$.
(b) Hall mobility $\mu_{\mbox{\tiny Hall}}$  {\em vs.} temperature
$T$ (Ref. \onlinecite{ohno}).
}
\end{figure}

\section{COMPARISON WITH EXPERIMENT}

First, we compare the SRT obtained from our
microscopic approach with the experimental data
by Ohno {\em et al.} in 60 periods of GaAs QWs separated by
$10-12$\ nm thick Al$_{0.4}$Ga$_{0.6}$As barriers.
The well width of each QW $a=7.5$\ nm and the electron
density $n=4 \times 10^{10}$\ cm$^{-2}$.\cite{ohno}
The well depth of the GaAs well confined by Al$_{x}$Ga$_{1-x}$As
is roughly estimated to be $65$\ \% of $1.087x+0.438x^2$\ eV,
for $x=0.4$ and is therefore  $V_0=328$\ meV.\cite{ot}
Differently from our previous fit\cite{wu7} with the experiment data
at high temperatures by Malinowski {\em et al.}\cite{harley2}
where we had two fitting parameters,
{\em i.e.}, the spin splitting parameter $\gamma$ and the impurity density
$n_i$ due to the absence of mobility data,
here we have only one  fitting parameter $\gamma$
and the temperature sweeps from the very low
temperature to the room temperature. The corresponding Hall mobilities
$\mu_{\mbox{\tiny Hall}}$ in the experiment\cite{ohno}
can be found in Ref.\ \onlinecite{kainz}, also plotted in Fig.\ 1(b).
From the Hall mobility,
one can deduce the impurity density by calculating the transport
mobilities\cite{lei} $\mu_{tr}=\mu_{\mbox{\tiny Hall}}/r_{\mbox{\tiny Hall}}$
with $r_{\mbox{\tiny Hall}}=1$ for the electron-AC phonon scattering due to the
deformation potential; $r_{\mbox{\tiny Hall}}=7/5$ for the electron-AC phonon
scattering due to the piezoelectric coupling and the electron-LO phonon
scattering; and $r_{\mbox{\tiny Hall}}=1$ for the
electron-ionized impurity scattering.\cite{rhall}

The only fitting parameter $\gamma$ is around
\begin{equation}
\gamma=(4/3)(m^{\ast}/m_{cv})(1/\sqrt{2m^{\ast
3}E_g})(\eta/\sqrt{1-\eta/3})\ ,
\label{gamma0}
\end{equation}
in which $\eta=\Delta/(E_g+\Delta)$;
$E_g$ denotes the band gap; $\Delta$ represents the spin-orbit
splitting of the valence band; $m^{\ast}$ stands for the electron mass
in GaAs; and $m_{cv}$ is a constant close in magnitude to the free
electron mass $m_0$.\cite{aronov} For GaAs when $m_{cv}=m_0$,
$\gamma=\gamma_0=11.4$\ eV$\cdot$\AA$^3$. The
initial spin polarization $P$
is assumed to be $2.5$\ \% for weak polarization throughout the paper.
In Fig.\ 1(a), the SRTs $\tau$ obtained from our approach
are plotted against the temperature with all the scattering included.
$B$ and $E$ are taken to be zero, as in the experiment.\cite{ohno}
$\gamma=0.9$, 1.0 and $1.1\gamma_0$ correspond
to $m_{cv}=1.1$, 1.0 and $0.91m_0$ with $m_{cv}$
being  the only not-fully-determined parameter  in
Eq.\ (\ref{gamma0}). One finds {\em good}
agreement between our theory and the experiment data almost over the {\em
whole} temperature regime.
When $T$ is below $13$\ K, there is no theoretical data due to
the lack of experimental data for the Hall mobility.
Kainz {\em et al.} also fitted the same experiment data
by using the single-particle theory without the Coulomb scattering.\cite{kainz}
They used a fourteen-band model to calculate the spin-orbit
coupling. Unlike our theory,
their results  can only give the
boundary values of the SRT in several cases
rather than the exact data.\cite{kainz}
This is because they did not take the full microscopic calculation,
and the single-particle theory is inadequate
in accounting for the spin R/D.

The best-fitted value of the spin-orbit coupling parameter, {\em i.e.},
$\gamma_0=11.4$\ eV$\cdot$\AA$^3$, is close to
the value calculated by Kainz {\em et al.}
$(\sim 16.5$\ eV$\cdot$\AA$^3)$ using the multiband
envelope-function approximation.\cite{kainz}
It is noted that the value for $\gamma$ in GaAs is still in debate.
Usually reported experimental values for $\gamma$
($25 - 30$\ eV$\cdot$\AA$^3$) in bulk material are deduced from the
DP spin relaxation mechanism within the framework of
the single-particle approximation, where the Coulomb scattering
is not included.\cite{marush}
Furthermore, the Raman scattering experiment showed that
$\gamma=16.5 \pm 3$ \ eV$\cdot$\AA$^3$ in asymmetric QW;\cite{juss}
and the Hanle effect experiment showed
that $\gamma=12.6$\ eV$\cdot$\AA$^3$.\cite{gore}
Theoretically, semi-empirical parameterized
$16 \times 16$ {\bf k}$\cdot${\bf p} calculations
show that $\gamma=14.9$\ eV$\cdot$\AA$^3$;\cite{cardona} and the
self-consistent {\em ab initio} calculations
predict $6.4$ and $8.5$\ eV$\cdot$\AA$^3$.\cite{chan}
Our fitting result supports the last four experimental and
theoretical results. It is also noted that our result further confirms
the analytical result Eq.\ (\ref{gamma0}) obtained from the
perturbation,\cite{aronov}  with $m_{cv}=m_0$.

\section{TEMPERATURE DEPENDENCE OF SRT}

We now study the temperature dependence of the spin relaxation in detail.
In the calculation, the electric field $E=0$, the magnetic field $B = 0$\ T and
the spin splitting parameter $\gamma = \gamma_0$.

We plot in Fig.\ 2 the temperature dependence of the SRT of
GaAs/Al$_{0.4}$Ga$_{0.6}$As QWs
with $a=7.5$\ nm at different impurity densities when the electron
densities are low ($n = 4 \times 10^{10}$\ cm$^{-2}$) [Fig.\ 2(a)],
medium ($n = 1 \times 10^{11}$\ cm$^{-2}$) [Fig.\ 2(b)] and high ($n
= 2 \times 10^{11}$\ cm$^{-2}$) [Fig.\ 2(c)] respectively as solid
curves. For the well width and the electron densities here, the
linear terms in the DP terms [Eqs.\ (\ref{omegax}-\ref{omegaz})] are
dominant, and only the lowest subband is relevant when $T\le300$\ K.
It is seen from the figure that adding impurities always increases
the SRT. This is understood that the criterion  of strong scattering
$|\mathbf{\Omega}|\tau_{p} \ll 1$ is satisfied here at all
temperatures, and therefore adding additional scattering always
increases the SRT.\cite{lue}  It is noted that $\tau_p$ here has
been extended to include $\tau_p^{ee}$, {\em i.e.}, the contribution
from the Coulomb scattering. \cite{zhushi1}
It is interesting to note that unlike our previous works focusing on
high temperatures ($T\ge120$\
K),\cite{wu4,wu5,wu6,wu7,wu8,wu9,wu10,lue,wu12} the situation is more
complicated at low temperatures. At low/medium electron densities
[Fig.\ 2(a)/(b)], the SRT presents a peak at very low temperature
(near 20$\sim$30\ K)/low temperature (around 41\ K) and a valley
around $120$\ K; whereas at high electron density [Fig.\ 2(c)],  the
SRT increases monotonically with $T$.

\begin{figure}[htb]
\psfig{figure=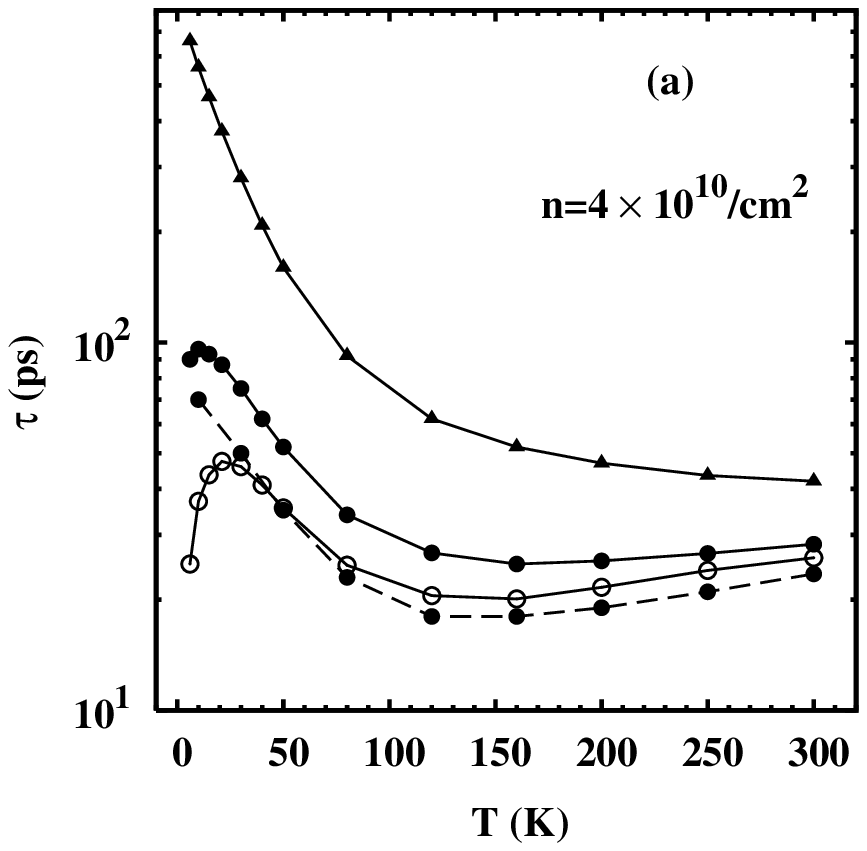,width=8cm,height=5cm,angle=0}
\psfig{figure=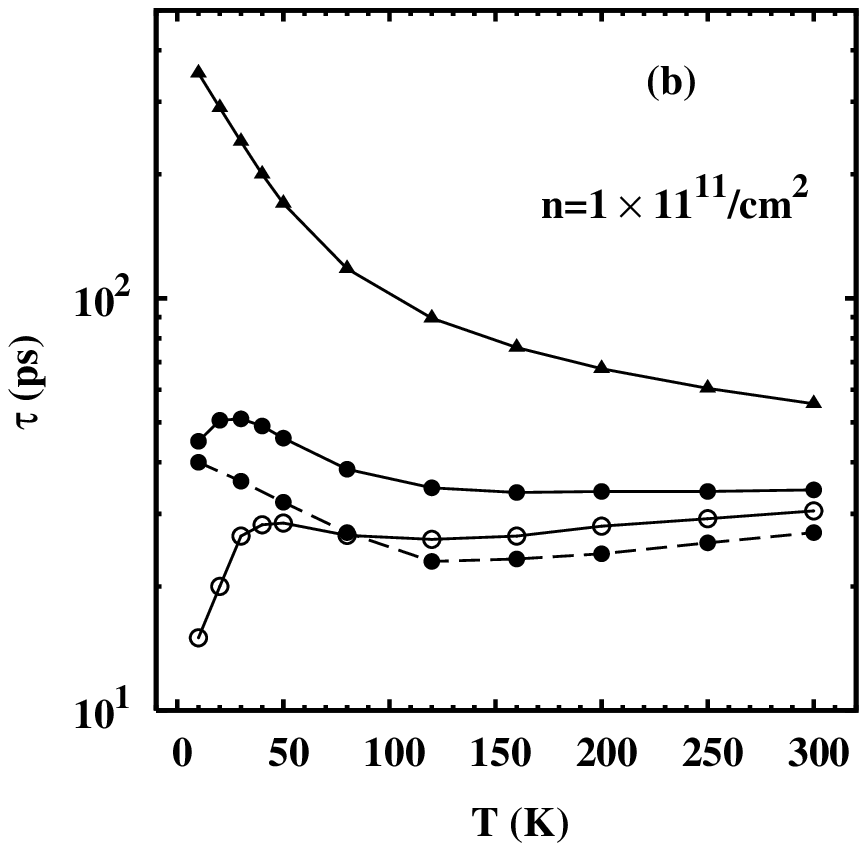,width=8cm,height=5cm,angle=0}
\psfig{figure=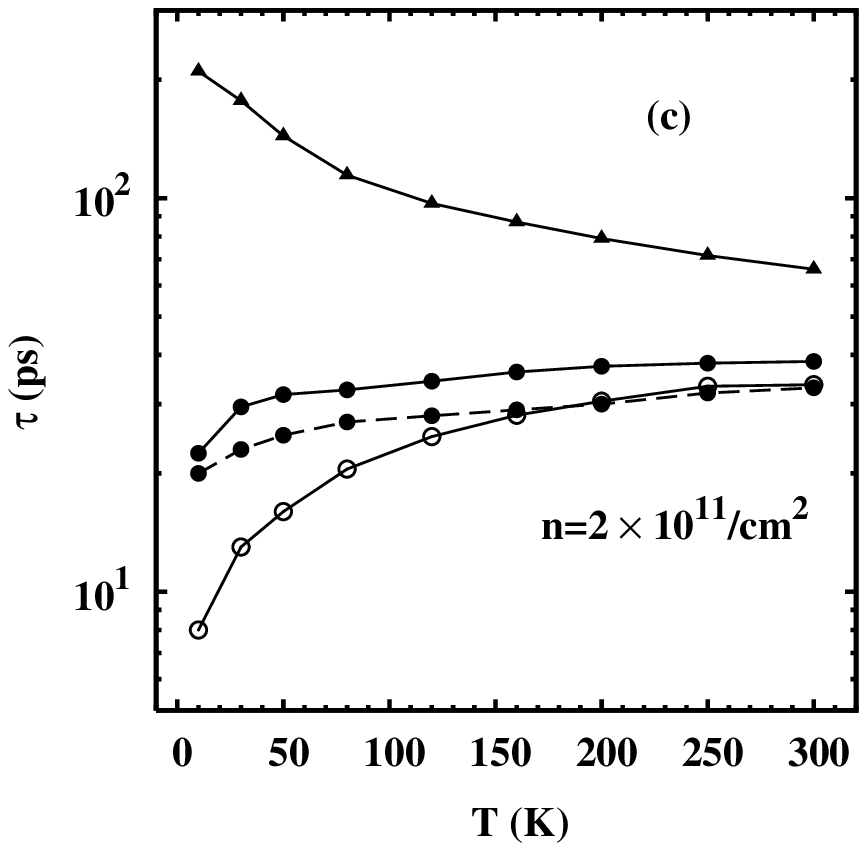,width=8cm,height=5cm,angle=0}
  \caption{SRT $\tau$ {\em vs} the temperature $T$
with  well width $a = 7.5$\ nm and electron density $n$ being
(a) $4 \times 10^{10}$\ cm$^{-2}$,
(b) $1 \times 10^{11}$\ cm$^{-2}$
 and (c) $2 \times 10^{11}$\ cm$^{-2}$ respectively.
Solid curve with triangles: $n_i = n$;
Solid curve with dots: $n_i = 0.1 n$;
Solid curve with circles:  $n_i = 0$;
Dashed curve with dots:  $n_i = 0.1 n$ and no Coulomb scattering.
}
\end{figure}

It is noted that at very low temperature (around 20\ K)
the electron-AC phonon scattering is negligible.\cite{yu}
$\tau_p^{AC}$ from the electron-AC phonon scattering is
around $25$\ ps,  two orders of magnitude
larger than  $\tau_{p}^{ee}$ from the electron-electron Coulomb scattering.
In addition, $\tau_p^i$ from the impurity scattering is
around 2\ ps, one order of magnitude larger than $\tau_{p}^{ee}$,
and has a very weak temperature dependence.
Therefore, the appearance of the peaks
in Fig.\ 2(a) originates from the electron-electron Coulomb scattering which
dominates the scattering process. Moreover, $\tau_{p}^{ee}$ is
a nonmonotonic function of temperature: $\tau_{p}^{ee} \propto T^{-2}$
at low temperature (degenerate limit) and  $\tau_{p}^{ee} \propto T$
at high temperature (nondegenerate limit).\cite{glazov}
The minimum of $\tau_{p}^{ee}$ corresponds to the crossover
from the degenerate limit to the  nondegenerate one
at $T_c \sim E_F/k_B$.
$T_c\sim 17$\ K when $n = 4 \times 10^{10}$\ cm$^{-2}$,
in good agreement with the peaks obtained from
our calculation with the exact Coulomb scattering.
Therefore the SRT increases/decreases with the temperature
(and the Coulomb scattering)
in the degenerate/nondegenerate regime.
Once $T \ge 120$\ K, the electron-LO phonon scattering
becomes comparable with the Coulomb scattering and
strengthens so rapidly with temperature that it completely
surpasses the weak temperature
dependence of the Coulomb scattering: $\tau_p^{LO}$ from
the electron-LO phonon scattering varies from several picoseconds at $120$\ K
to several tenths of picosecond at $300$\ K,
and $\tau_{p}^{ee}$ varies from $1$\ ps
to several picoseconds.
Therefore the SRT increases with $T$.
When the electron density is $ 1 \times 10^{11}$\ cm$^{-2}$,
$T_c$ is nearly $41$\ K. Around this temperature, the electron-AC
phonon scattering cannot be overlooked although
$\tau_p^{AC}$ is still roughly one order of magnitude larger than
$\tau_{p}^{ee}$. Therefore, the reduction of the Coulomb scattering
after $T_c$ can be partly compensated by the increase of the
electron-phonon scattering. As a result,
one can see that the decrease of the SRT after $T_c$ in Fig.\ 2(b)
is much slower than that in Fig.\ 2(a). However,
when the electron density is high enough, say $2\times 10^{11}$\ cm$^{-2}$
 in Fig.\ 2(c), $T_c$ is nearly $83$\ K, much larger
than the case of low electron density. At this temperature,
the electron-phonon scattering becomes comparable to the Coulomb
scattering and the strengthening rate of phonon scattering around this
temperature is large enough to completely compensate, and even surpass, the
weakening rate of the Coulomb scattering.
Consequently the total scattering
increases monotonously with $T$. Therefore the SRT increases
monotonically with $T$.

We further show the effect of the  Coulomb scattering on the
spin relaxation. This was first proposed
by Wu and Ning based on the inhomogeneous broadening induced by the
energy dependence of the $g$-factor.\cite{wu2}
Then, we used our full microscopic approach and
showed that the Coulomb scattering makes marked contribution to
the spin R/D when $T\ge120$\ K when the inhomogeneous
broadening is induced by the DP term.\cite{wu7,wu8,lue}
At low temperature ($T<120$\ K), Glazov and Ivchenko used perturbation method
to show that the second-order Coulomb
scattering causes the SRT.\cite{glazov1} In the perturbation approach, the
Coulomb scattering contributes marginally to the spin R/D at high temperature.
In our calculation, we include the Coulomb scattering to all orders
of the bubble diagrams as
well as the counter effect of the Coulomb scattering to the inhomogeneous
broadening.  In Fig.\ 2 by plotting the SRT for the case of  $n_i = 0.1n$, but
without the Coulomb scattering, as dashed curves, we show that the
Coulomb scattering makes marked contribution to the spin R/D over the
whole temperature regime by increasing the spin R/D time.\cite{comment}
It is further seen from Figs.\ 2(a) and (b)
that the peak disappears without the
Coulomb scattering. This is consistent with the previous discussion.

It is interesting to see that in the absence of the Coulomb
scattering, the criterion for  strong scattering regime
$|\mathbf{\Omega}|\tau_{p} \ll 1$
is satisfied only when $T > 120$\ K. Therefore the SRT increases with $T$
when $T\ge120$\ K.
When $T < 120$\ K, $|\mathbf{\Omega}|\tau_{p}$ is slightly smaller than $1$,
which is the intermediate scattering regime.
The variation of the SRT depends on
the competition between the increase of the inhomogeneous
broadening and the increase of the scattering with the temperature.\cite{lue}
For low/high electron density case,
the temperature dependence of the electron-AC phonon scattering is
less/more effective and the SRT decreases/increases with $T$.

\begin{figure}[htb]
\psfig{figure=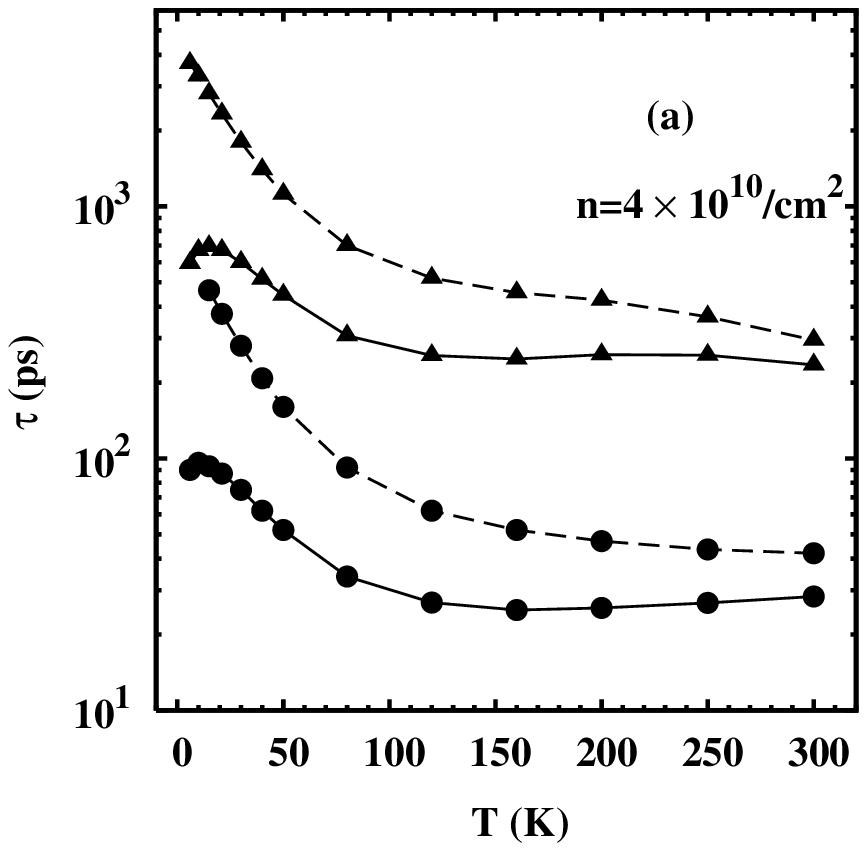,width=8cm,height=5cm,angle=0}
\psfig{figure=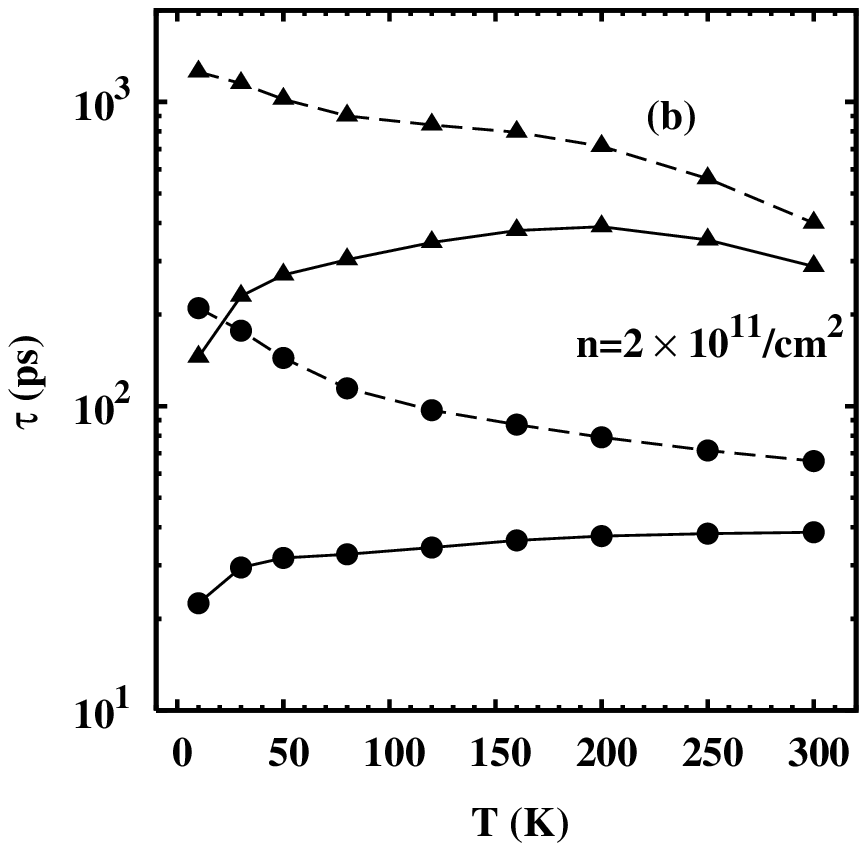,width=8cm,height=5cm,angle=0}
\caption{SRT $\tau$ {\em vs.} temprature T at
$a = 7.5$\ nm (curves with dots) and 15\ nm (curves with
triangles). Solid curves: $n_i = 0.1 n$; Dashed curves:
$n_i = n$. (a)
$n = 4 \times 10^{10}$\ cm$^{-2}$ and
(b)  $n = 2 \times 10^{11}$\ cm$^{-2}$.}
\end{figure}

It is noted that in order to see the peaks at low electron density,
it is important to have a high mobility sample (low impurity density).
This is because the ascendancy of the Coulomb scattering
can be impaired when the impurity scattering gets large enough and
the total scattering is mainly determined by the impurity scattering.
As the electron-impurity scattering depends weakly on the temperature,
the temperature dependence of the inhomogeneous broadening
from the DP term becomes the only variable element.
Therefore the SRT decreases monotonically with $T$ as the solid curves
with triangles in Fig.\ 2  for the case of $n_i = n$.
This condition is not satisfied in the
experiment by Ohno {\em et al.},\cite{ohno}
and this is the reason why there is no peak in Fig.\ 1. However, apart
from the peaks which are not observed yet,
both the experiment and calculation show that
the SRT decreases with temperature at low electron densities
when $T<120$\ K. The  SRT at high electron density increases monotonically
with temperature when the impurity density
is low, which is also in agreement with the latest experiment
by Harley {\em et al.}\cite{harley3}

Finally we investigate the well width dependence of the SRT.
In Fig.\ 3 we plot the SRT versus temperature at
well widths $a=7.5$\ nm (solid curves) and 15\ nm (dashed
curves) respectively. We choose low and high impurity densities
$n_i=0.1n$  (curves with dots) and $n_i=n$ (curves
with triangles) as well as low and high electron densities
$n = 4 \times 10^{10}$\ cm$^{-2}$ (a) and
$n = 2 \times 10^{11}$\ cm$^{-2}$ (b).
It is noted that the SRT is enhanced by increasing the well width
as   $\langle k_z^2 \rangle$ in
the DP term decreases with the increase of  $a$. Moreover, as impurities
further enhance the SRT, it reaches to several nanoseconds at
very low temperatures at high impurity density.

\section{Electric field dependence of SRT}

We now turn to investigating the hot-electron effect on spin relaxation at
low temperature. An electric field is applied parallel to the QW.
Similar to our previous investigation,\cite{wu8} electrons obtain a
center-of-mass drift velocity (and consequently an effective
magnetic field proportional to the electric field) and
are heated to a temperature $T_e$ higher than $T$.
The numerical schemes of solving the hot electron problem
has been  laid out in detail in Ref.\ \onlinecite{wu8}.\cite{comment1}
We plot the electric field dependence of the
SRT with $a = 7.5$\ nm and $T=50$\ K
for different impurity densities at low/high electron density
($n = 4 \times 10^{10}$\ cm$^{-2}$/$n = 2 \times 10^{11}$\ cm$^{-2}$)
in Fig.\ 4(a)/(b). In the calculation, the magnetic field
$B = 4$\ T and the spin splitting parameter
$\gamma = \gamma_0$ as in the previous section.
It is seen from the figure that unlike the high temperature case
investigated before\cite{wu8}
(and also see Fig.\ 4(c) for $T=120$\ K)
where the electric field can be applied easily to around 1\ kV/cm,
at low temperatures it can be applied only to a very small
value due to the ``runaway'' effect.\cite{run} This is because
at low temperature, the efficient electron-LO phonon scattering is
missing and electrons are therefore very easily driven to very high
momentum states by a very small electric field.

It is interesting to see from the figure that differing from
the high temperature case where the SRT {\em increases}
 with the electric field (see
Fig.\ 4 (c) and also Ref.\ \onlinecite{wu8}),
here, for the case of low electron densities,
the SRT {\em decreases} with the field and for the case of high
electron densities, the SRT decreases/increases with the field
at high/low impurity densities.

\begin{figure}[htb]
\centerline{\psfig{figure=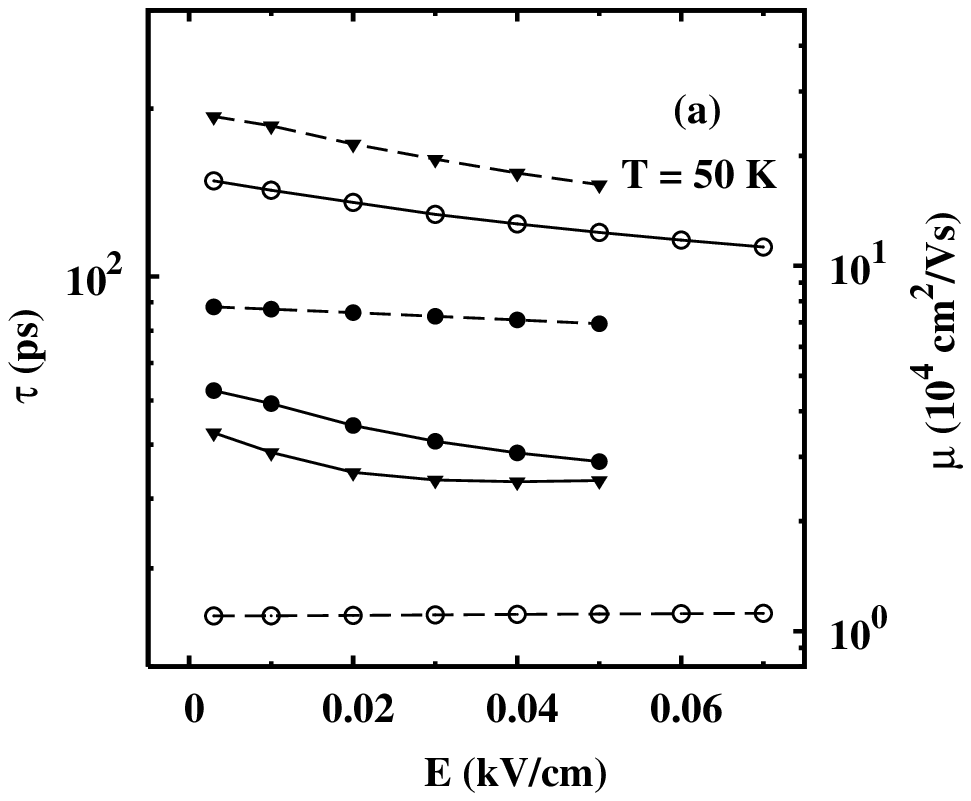,width=8cm,height=5cm,angle=0}}
\centerline{\psfig{figure=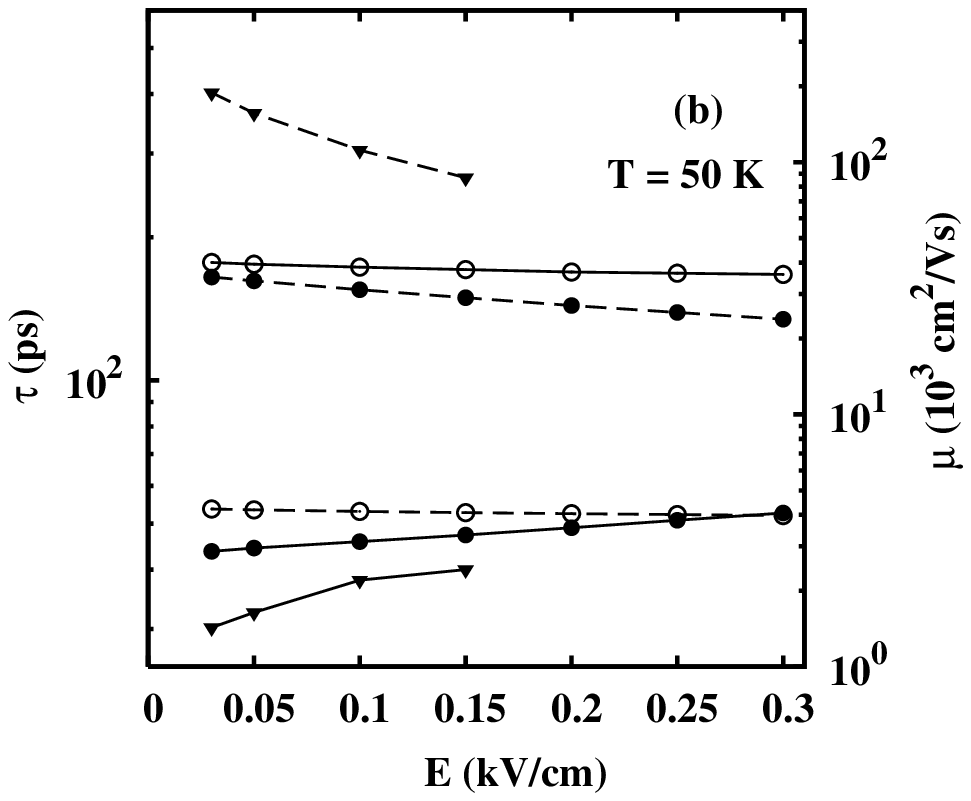,width=8cm,height=5cm,angle=0}}
\centerline{\psfig{figure=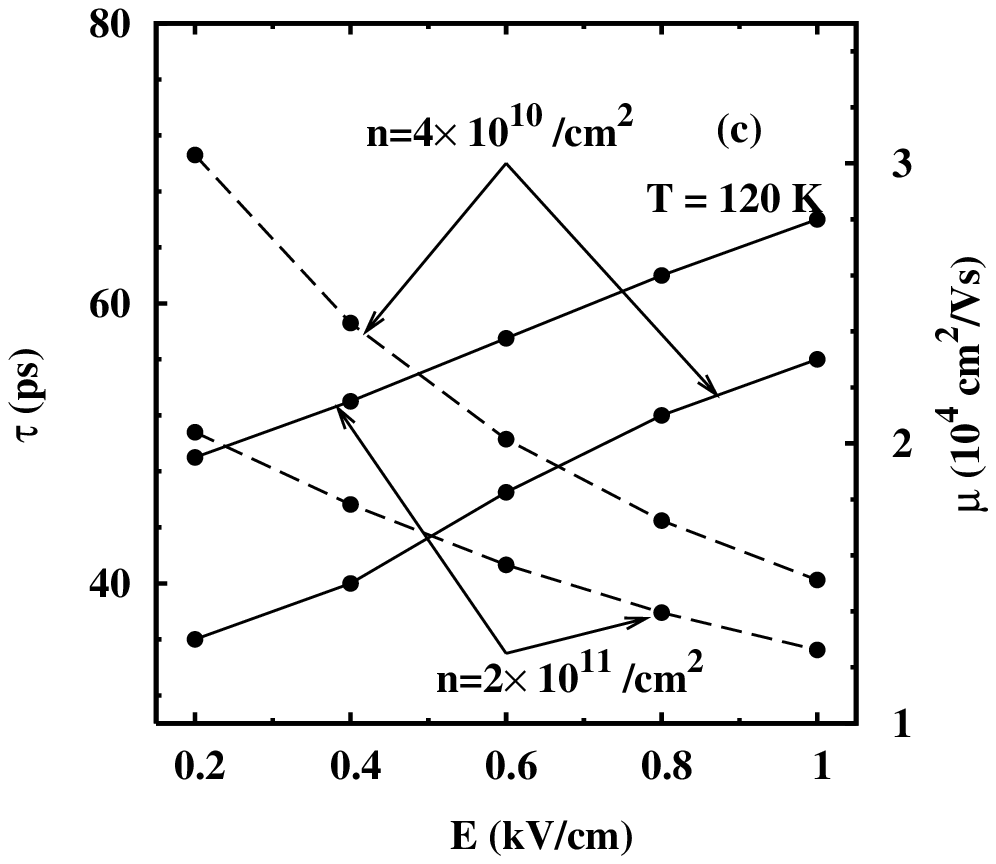,width=8cm,height=5cm,angle=0}}
\caption{SRT $\tau$ (solid curves) and mobility
$\mu$ (dashed curves) {\em vs.} electric field $E$.
(a) $T = 50$\ K and $n = 4 \times 10^{10}$\ cm$^{-2}$;
(b) $T=50$\ K and  $n = 2 \times 11^{11}$\ cm$^{-2}$;
(c) $T=120$\ K, $n = 4 \times 10^{10}$\ cm$^{-2}$ and
$2 \times 11^{11}$\ cm$^{-2}$ respectively.
Curves with open circles: $n_i = n$; with
dots: $n_i = 0.1 n$; with triangles: $n_i = 0$.
Note the scales of the mobility $\mu$ are on the right side of the figures.
}
\end{figure}

\begin{figure}[htb]
\centerline{\psfig{figure=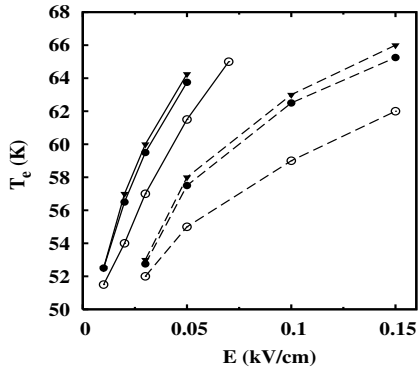,width=8cm,height=5cm,angle=0}}
\caption{Hot electron temperature $T_e$ {\em vs.} electric field $E$ when
$T = 50$\ K. solid curves:  $n = 4 \times 10^{10}$\ cm$^{-2}$;
dashed curves: $n = 2 \times 11^{11}$\ cm$^{-2}$.
Curves with open circles: $n_i = n$; with dots: $n_i = 0.1 n$;
with triangle: $n_i = 0$.}
\end{figure}

These features at low temperature $T$ can be understood from the joint
effects of the electric field $E$ to the scattering strength
and the inhomogeneous broadening due to the DP term.
On one hand, when the electric field is small,
the ionized-impurity scattering, whose strength decreases slightly
with the electron temperature $T_e$, is dominant.
When the electric field is further increased, $T_e$ and
therefore the electron-AC phonon scattering is raised.
If the impurity scattering is not too high, the
electron-AC phonon scattering can then be
dominant as discussed decades ago in Ref.\ \onlinecite{conwell}.
These can be seen from the mobilities $\mu=\sum_{{\bf k}\sigma}
\hbar{\bf k}f_{{\bf k}\sigma}/(m^\ast nE)$ obtained from our
calculation, which are
plotted in the same figure for all the corresponding cases.
One can see that $\mu$ increases slightly and monotonically with $E$ for the
impurity-scattering-dominant case such as $n_i = n$;
It decreases monotonically with $E$ for the case of very
low/no impurity scattering such as
$n_i = 0$, as the electron-AC phonon scattering always increases
 with the electron temperature $T_e$;
for the case of low impurity scattering such as $n_i = 0.1 n$,
$\mu$ first increases slightly then decreases with $E$ which shows the
transition from the impurity-scattering-dominant regime to
the electron-AC-phonon-scattering-dominant regime\cite{conwell} (unless
the runaway effect blocks the system to the later regime as shown in
Fig.\ 4(a) for the case of low electron densities).
On the other hand, electrons are driven to the higher momentum states
by the electric field and  experience a larger effective magnetic field
[Eqs.\ (\ref{omegax}-\ref{omegaz})]. Therefore the inhomogeneous broadening
is increased. This tends to decrease the SRT.
In order to show the electric field dependence
of the inhomogeneous broadening,
we plot in Fig.\ 5 the electron temperature
$T_e$ as a function of electric field
$E$ for all the corresponding cases in Figs.\ 4(a) and (b).
It is obvious from Fig.\ 5 that the increase of $T_e$ for the
low electron density case is much faster than that
for the high one. Therefore, the increase of the inhomogeneous broadening is
the leading contribution in comparison with the electric field effect on
the scattering. Consequently, the SRT decreases with $E$ for
the case of low electron densities as shown in Fig.\ 4(a).
For the case of high electron densities,
when the impurity density is high such as $n_i=n$ in Fig.\ 4(b),
both the slight decrease of the scattering and the increase of
the inhomogeneous broadening tend to suppress the SRT.
When the impurity density is  low/zero,
the strengthening of the scattering is dominant
(as shown in Fig.\ 4(b) the decrease of
mobility with $E$) in comparison with the increase of the inhomogeneous
broadening. This makes the SRT decrease with $E$.
Finally we point out that the Coulomb scattering plays an essential
role in the spin R/D in the presence of the electric field. It determines
the hot-electron temperature $T_e$ which controls the inhomogeneous
broadening and the scattering strengths.
Moreover, the Coulomb scattering itself
also contributes to the spin R/D.

For comparison, we also plot the SRT versus electric field
at high temperature  $T = 120$\ K for both low and high electron densities
with $n_i=0.1n$ in Fig.\ 4(c).
At this temperature, the electron-LO phonon scattering is dominant.
Therefore $\mu$ always decrease with $E$.
For the well width and electron density we study,
the linear DP term is dominant and the increase of the scattering is
more important. Therefore the SRT increases with $E$. For QWs with
larger well width so that the cubic term is dominant, the SRT can
decrease with $E$ as reported in our previous work at high
temperatures.\cite{wu9}

\section{Summary}

In summary, we have investigated the
temperature dependence of the SRT for $n$-type GaAs (001)
QWs with small well widths from a full microscopic approach
by constructing and numerically
solving the kinetic spin Bloch equations with all the 
relevant  scattering
explicitly included. In contrast to our previous studies at
high temperatures ($T\ge 120$\ K),
we include the electron-AC phonon scattering which is absent
in our previous studies so that we may extend the scope of our approach
to the low temperature regime ($T < 120$ K).
Good agreement with experiment data\cite{ohno} is obtained
from our theory  over almost the {\em whole} temperature regime
by using only one fitting parameter $\gamma$ whose value
agrees with many experimental and theoretical results.
We  show that the Coulomb scattering plays an essential
role in spin R/D over all the temperature regime.

For QWs with low electron densities but high mobility ({\em i.e.},
low impurity density), the spin R/D
is mainly controlled by the electron-electron Coulomb scattering when
$T< 70$\ K. We predict a peak in the $\tau$-$T$ curve. The closer the
peak approaches the high temperature limit, the smoother the peak appears.
After the peak, the SRT increases with temperature.
Finally, such a peak disappears at sufficient high electron density
where the SRT increases monotonically
with temperature. We point out that the
peak origins from the Coulomb scattering. Specifically, it origins from
the different temperature dependences of the Coulomb
scattering at the degenerate and the nondegenerate limits with
the transition temperature $T_c\sim E_F/k_B$. For low electron densities,
$T_c\le 30$\ K where the electron-phonon scattering is negligible. Then
one may observe an abrupt peak around $T_c$.
For medium electron densities,
$30$\ K$ < T_c < 70$\ K where the increase of the electron-AC phonon
scattering partially compensates the decrease
of the Coulomb scattering when $T$ increases,
one may observe a smooth peak around $T_c$.
Nevertheless, for high electron
densities, $T_c>70$\ K, the increase of the electron-phonon scattering
completely compensates the decrease of the Coulomb scattering when $T$ rises.
Consequently the peak disappears.

At high temperature ($T\ge120$\ K) and low impurity density,
when the well width is small so that the cubic terms in the DP terms
are unimportant, the increase of electron-LO phonon scattering surpasses
the increase of inhomogeneous broadening with temperature, so that the SRT
increases with temperature. However,
when the impurity density is so high that
electron-impurity scattering is the dominant
scattering mechanism, the SRT decreases monotonically with
temperature for any electron density. This is because the temperature
dependence of the electron-impurity scattering is very weak and the increase
of the inhomogeneous broadening with temperature dominates
the temperature dependence of the SRT.
We also show that larger well width  leads to a slower spin relaxation.
Moreover, in the strong scattering limit, higher impurity density
also leads to a slower spin relaxation. Both effects can make the SRT
as long as nanoseconds at very low temperatures.

The effect of electric field ({\em i.e.}, the hot electron effect) on
the spin relaxation is also investigated. We
show that the electric field dependence of the SRT at low temperature
appears again quite differently from that at high temperature due to the
absence of electron-LO phonon scattering. Moreover, we further show
different electric field dependences of the SRT  at low and high electron densities.
At low electron densities, the SRT {\em decreases} with the electric field.
When the electron density is high, it {\em decreases/increases}
 with the electric field for the case of high/low impurity densities.
These features are in giant difference from the high temperature case where
the SRT {\em increases} monotonically with electric field for the same QWs.
More experiments are needed to explore the predictions presented in
this manuscript.

\begin{acknowledgments}

This work was supported by the Natural Science Foundation of China
under Grant Nos.\ 90303012 and 10574120, the Natural Science Foundation
of Anhui Province under Grant No.\ 050460203, the Knowledge Innovation
Project of Chinese Academy of Sciences and SRFDP. The authors would like to 
thank I. C. da Cunha Lima for his critical reading of this manuscript to 
improve the English.

\end{acknowledgments}

\begin{appendix}
\begin{section}{Numerical scheme for electron-AC phonon scattering}

The electron-AC phonon scattering terms can be rewritten as
\begin{widetext}
\begin{eqnarray}
\left.\frac{\partial f_{{\bf k},\sigma}}{\partial t}\right|_{\mbox{\small
AC}}&=&\Bigl\{-2\pi\sum _{{\bf q}q_{z},\lambda}
g^2_{{\bf q} q_{z},\lambda}\delta
(\epsilon_{\bf k}-\epsilon_{{\bf k}-{\bf q}}-
\Omega_{{\bf q} q_{z}\lambda})  
[N(\epsilon_{\bf k}-\epsilon_{{\bf k}-{\bf q}})
(f_{{\bf k},\sigma}-f_{{\bf k}-{\bf q},\sigma})
+f_{{\bf k},\sigma}(1-f_{{\bf k}-{\bf q},\sigma})\nonumber\\
&&-\mbox{Re}({\rho}_{{\bf k}} {\rho}^{\ast}_{{\bf k-q}})]\Bigr\}
-\Bigl\{{\bf k} \leftrightarrow {\bf k-q}\Bigr\}\ ,\\
\left. \frac{\partial \rho_{\mathbf{k}}}{\partial t}\right |_{\mbox{\small
AC}}&=&\Bigl\{
  \pi\sum_{\mathbf{q}q_z\lambda}g^2_{\mathbf{q}q_z\lambda}
  \delta(\varepsilon_{\mathbf{k}}-\varepsilon_{\mathbf{k}-\mathbf{q}}
  -\Omega_{\mathbf{q}q_z\lambda})
[\rho_{\mathbf{k}-\mathbf{q}}
  (f_{\mathbf{k},\uparrow}+f_{\mathbf{k},\downarrow})
  +(f_{\mathbf{k}-\mathbf{q},\uparrow}+f_{\mathbf{k}-\mathbf{q},\downarrow}-2)
  \rho_{\mathbf{k}}\nonumber\\
&&-2N(\epsilon_{\bf k}-\epsilon_{{\bf k}-{\bf q}})
  (\rho_{\mathbf{k}}-\rho_{\mathbf{k}-\mathbf{q}})]
\Bigr\} -\Bigl\{\mathbf{k}\leftrightarrow \mathbf{k}-\mathbf{q}\Bigr\}\ ,
\end{eqnarray}
\end{widetext}
with $\rho_{{\bf k}}\equiv \rho_{{\bf k},\uparrow\downarrow}$
and $\{{\bf k}\leftrightarrow {\bf k}-{\bf q}\}$ standing for the
same terms as the previous $\{\}$ but with the interchange
${\bf k}\leftrightarrow {\bf k}-{\bf q}$. $N(\epsilon_{\bf k}
-\epsilon_{{\bf k}-{\bf q}})=[\exp\{\beta(\epsilon_{\bf k}
-\epsilon_{{\bf k}-{\bf q}})\}-1]^{-1}$  represents the Bose distribution.
The division of the  truncated two-dimensional momentum space
is all the same as our previous work (see Fig.\ 8 in Ref.\ \onlinecite{wu8}).
The two dimensional momentum space is thus divided into $N\times M$
control regions, each with the same energy and  angle intervals.
The ${\bf k}$-grid point of each control region is chosen to be
the center of the region:
\begin{equation}
\mathbf{k}_{n,m}=\sqrt{2m^{\ast}E_n}(\cos \theta_{m},\sin \theta_{m}),
\end{equation}
with $E_{n}=(n+1/2)\Delta_E$ and $\theta_{m}=m\Delta \theta$.
$n=0,1,\cdots,N-1$ and $m=0,1,\cdots,M-1$.
with the truncation energy $E_{cut} = E_{N}$
and $\theta_{M}=(M-1)2\pi/M$.

Unlike the electron-LO phonon scattering where the $\delta$-function
in the scattering is used to carry out the integral of ${\bf k}^\prime$,
more specifically  $\theta_{{\bf k}^{\prime}}$, with
${\bf k}^\prime\equiv {\bf k}-{\bf q}$, here the
$\delta$-function is used to perform the integral of $q_z$ with
\begin{equation}
q_z=\sqrt{(\frac{\epsilon_{{\bf k}}
-\epsilon_{{\bf k}^{\prime}}}{v_{\lambda}})^2-{\bf q}^2}\ .
\end{equation}

\end{section}
\end{appendix}


\begin{thebibliography}{10}
\bibitem{meier} {\em Optical Orientation}, edited by F. Meier and B. P. Zakharchenya, (North-Holland, Amsterdam, 1984).

\bibitem{prinz}
{\em Semiconductor Spintronics and Quantum
  Computation}, eds. D. D. Awschalom, D. Loss, and N. Samarth
  (Springer, Berlin, 2002); I. \v Zuti\'c, J. Fabian, and S. Das Sarma,
Rev. Mod. Phys. {\bf 76}, 323 (2004).
\bibitem{kikka}J. M. Kikkawa and D. D. Awschalom, Phys. Rev. Lett. {\bf 80}, 4313 (1998).
\bibitem{dzhioev} R. I. Dzhioev, B. P. Zakharchenya, V. L. Korenev, D. Gammon, and D. S. Katzer, Pis'ma  Zh. \'{E}ksp. Teor. Fiz. {\bf 74}, 200 (2001)
[JETP Lett. {\bf 74}, 182 (2001)];
R. I. Dzhioev, K. V. Kavokin, M. V. Lazarev, B. Ya. Meltser, M. N. Stepanova,
 B. P. Zakharchenya, D. Gammon, and D. S. Katzer,
Phys. Rev. B {\bf 66}, 245204 (2002).

\bibitem{murdin} B. N. Murdin, K. Litvinenko, J. Allam, C. R. Pidgeon, M. Bird, K. Morrison, T. Zhang,
  S. K. Clowes, W. R. Branford, J. Harries, and L. F. Cohen, Phys. Rev. B {\bf 72}, 085346 (2005).

\bibitem{damen} T. C. Damen, L. Vi\~{n}a, J. E. Cunningham, J. Shah, and L. J. Sham, Phys. Rev. Lett. {\bf 67},
  3432 (1991).
\bibitem{wagner} J. Wagner, H. Schneider, D. Richards, A. Fischer, and K. Ploog, Phys. Rev. B {\bf 47}, 4786 (1993).
\bibitem{heberle} A. P. Heberle, W. W. R\"{u}hle, and K. Ploog, Phys. Rev. Lett. {\bf 72}, 3887 (1994).

\bibitem{crooker} S. A. Crooker, J. J. Baumberg, F. Flack, N. Samarth, and D. D. Awschalom,
  Phys. Rev. Lett. {\bf 77}, 2814 (1996); S. A. Crooker, D. D. Awschalom, J. J. Baumberg,
F. Flack, and N. Samarth, Phys. Rev. B {\bf 56}, 7574 (1997).

\bibitem{kikka1} J. M. Kikkawa, I. P. Smorchkova, N. Samarth, 
and D. D. Awschalom, Science {\bf 277}, 1284
  (1997).
\bibitem{ohno2} Y. Ohno, R. Terauchi, T. Adachi, F. Matsukura, and H. Ohno, Phys. Rev. Lett. {\bf 83}, 4196
  (1999).
\bibitem{ohno} Y. Ohno, R. Terauchi, T. Adachi, F. Matsukura, and H. Ohno,
Physica E (Amsterdam) {\bf 6}, 817 (2000).
\bibitem{harley2} A. Malinowski, R. S. Britton, T. Grevatt, R. T. Harley, D. A. Ritchie, and M. Y. Simmons,
  Phys. Rev. B {\bf 62}, 13034 (2000); M. A. Brand,
 A. Malinowski, O. Z. Karimov, P. A. Marsden, R. T. Harley,
  A. J. Shields, D. Sanvitto, D. A. Ritchie, and M. Y. Simmons, Phys. Rev. Lett. {\bf 89}, 236601 (2002).
\bibitem{ohno1} T. Adachi, Y. Ohno, F. Matsukura, and H. Ohno, Physica E (Amsterdam) {\bf 10}, 36 (2001).
\bibitem{harley1} O. Z. Karimov, G. H. John, R. T. Harley, W. H. Lau, M. E. Flatt\'{e}, M. Henini, and  R. Airey, Phys. Rev. Lett. {\bf 91}, 246601 (2003).
\bibitem{doh} S. D\"{o}hrmann, D. H\"{a}gele, J. Rudolph, M. Bichler, D. Schuh, and M. Oestreich,  Phys. Rev. Lett. {\bf 93}, 147405 (2004).
\bibitem{lombez} L. Lombez, P. -F. Braun, H. Carr\`{e}re, B. Urbaszek, P. Renucci. T. Amand, X. Marie, J. C. Harmand, and V. K. Kalevich,
Appl. Phys. Lett. {\bf 87}, 252115 (2005).
\bibitem{strand} J. Strand, X. Lou, C. Adelmann, B. D. Schultz, A. F. Isakovic, C. J. Palmstr$\phi$m, and
  P. A. Crowell, Phys. Rev. B {\bf 72}, 155308 (2005).
\bibitem{shaff}A. M. Tyryshkin, S. A. Lyon, W. Jantsch, and F. Sch\"affler,
Phys. Rev. Lett. {\bf 94}, 126802 (2005).

\bibitem{holl} A. W. Holleitner, V. Sih, R. C. Myers, A. C. Gossard, and D. D. Awschalom, cond-mat/0602155.

\bibitem{dp} M. I. D'yakonov and V. I. Perel', Zh. \'{E}ksp. Teor. Fiz. {\bf 60} 1954 (1971). [Sov. Phys. JEPT {\bf 33}, 1053 (1971)].

\bibitem{dress} G. Dresselhaus, Phys. Rev. {\bf 100}, 580 (1955).

\bibitem{rashba} Y. A. Bychkov and E. I. Rashba, Pis'ma Zh. \'{E}ksp. Teor. Fiz. {\bf 39}, 66 (1984)
  [Sov. Phys. JEPT Lett. {\bf 39} 78 (1984)].

\bibitem{dp2}D'yakonov and Kachorovskii, Fiz. Tekh. Poluprovodn. v 20, p 178 (1986)
[Sov. Phys. Semicond. {\bf 20}, 110 (1986)].
\bibitem{car} X. Cartorix\`{a}, L. -W. Wang, D. Z. -Y. Ting, Y. -C. Chang, Phys. Rev. B {\bf 73}, 205341 (2006).
\bibitem{lau} W. H. Lau and M. E. Flatt\'e, Phys. Rev. B {\bf 72},
161311(R) (2005).

\bibitem{kainz} J. Kainz, U. R\"{o}ssler, and R. Winkler, Phys. Rev. B {\bf 70}, 195322 (2004).

\bibitem{single} W. H. Lau, J. T. Olesberg, and M. E. Flatt\'{e}, Phys. Rev. B {\bf 64}, 161301(R)
  (2001); P. H. Song and K. W. Kim, Phys. Rev. B {\bf 66}, 035207 (2002);
F. X. Bronold, I. Martin, A. Saxena,
  and D. L. Smith, Phys. Rev. B {\bf 66}, 233206 (2002);
N. S. Averkiev, L. E. Golub, and M. Willander,
J. Phys.: Condens. Matter {\bf 14}, R271 (2002); S. Krishnamurthy,
 M. van Schilfgaarde, and N. Newman,
  Appl. Phys. Lett. {\bf 83}, 1761
(2003);  F. X. Bronold, A. Saxena, and D. L. Smith, Phys. Rev. B
{\bf 70}, 245210 (2004); S. W. Chang and S. L. Chuang,
Phys. Rev. B {\bf 72}, 115429
(2005); Z. G. Yu, S. Krishnamurthy, M. van Schilfgaarde,
and N. Newman, Phys. Rev. B {\bf 71}, 245312 (2005);
O. Bleibaum, Phys. Rev. B {\bf 69}, 205202 (2004); {\bf 71}, 235318 (2005);
X. Cartoix\`a, D. Z. Y. Ting, and Y. C. Chang, Phys. Rev. B {\bf 71},
045313 (2005).

\bibitem{wu1} M. W. Wu and H. Metiu, Phys. Rev. B {\bf 61}, 2945 (2000);
M. W. Wu, J. Supercond.  {\bf 14}, 245 (2001).
\bibitem{wu2} M. W. Wu and C. Z. Ning, Eur. Phys. J. B {\bf 18}, 373 (2000).
\bibitem{wu3} M. W. Wu, J. Phys. Soc. Jpn. {\bf 70}, 2195 (2001).
\bibitem{wu4} M. W. Wu and C. Z. Ning, Phys. Stat. Sol.(b) {\bf 222},
 523 (2000);  M. W. Wu and M. Kuwata-Gonokami, Solid State Commun.
{\bf 121}, 509 (2002).
\bibitem{wu5} J. L. Cheng, M. Q. Weng, and M. W. Wu,  Solid State Commun.
{\bf  128}, 365 (2003).
\bibitem{wu6} M. Q. Weng and M. W. Wu, Phys. Rev. B {\bf 66}, 235109 (2002);
J. Appl. Phys. {\bf 93}, 410 (2003); M. Q. Weng, M. W. Wu, and
Q. W. Shi, Phys. Rev. B {\bf 69}, 125310 (2004); L. Jiang, M. Q. Weng,
M. W. Wu, and J. L. Cheng, J. Appl. Phys. {\bf 98}, 113702 (2005).
\bibitem{wu7} M. Q. Weng and M. W. Wu, Phys. Rev. B {\bf 68}, 075312 (2003);
{\bf 71}, 199902(E) (2005); Chin. Phys. Lett. {\bf 22}, 671 (2005).
\bibitem{wu8} M. Q. Weng, M. W. Wu, and L. Jiang, Phys. Rev. B {\bf 69}, 245320 (2004).
\bibitem{wu9} M. Q. Weng and M. W. Wu, Phys. Rev. B {\bf 70}, 195318 (2004).
\bibitem{wu10} L. Jiang and M. W. Wu, Phys. Rev. B {\bf 72}, 033311 (2005).
\bibitem{lue} C. L\"{u}, J. L. Cheng, and M. W. Wu, Phys. Rev. B
{\bf 73}, 125314 (2006).
\bibitem{wu12} J. L. Cheng and M. W. Wu, J. Appl. Phys. {\bf 99}, 083704 (2006).

\bibitem{ya} E. Ya. Sherman, Appl. Phys. Lett. {\bf 82}, 209 (2003).

\bibitem{haug} H. Haug and A. P. Jauho, {\em Quantum Kinetics in
Transport and Optics of Semiconductor} (Spinger-Verlag, Berlin, 1996).
\bibitem{vogl} P. Vogl, in  {\em Physics of Nonlinear
Transport in Semiconductor}, edited by K. Ferry, J. R. Barker, and
  C. Jacoboni  (Plenum, New York, 1980).
\bibitem{mahan}
G. D. Mahan, in
{\em Polarons in Ionic Crystals and Polar Semiconductor}, edited by J.
T. Devreese (North-Holland, Amsterdam, 1972).
\bibitem{land} {\em Semiconductors}, Landolt-B\"ornstein,
New Serious, Vol.\ 17a, ed. by O. Madelung (Springer, Berlin, 1987).
\bibitem{RPA} It is noted that the random phase approximation is valid
for all the electron densities we discuss
in the present paper according to the criterion given by
M. Jonson, J. Phys. C {\bf 9}, 3055 (1976).
\bibitem{haug1}H. Haug and S. W. Koch, {\em Quantum theory of the
optical and electronic properties of semiconductors} (World Scientific,
Singapore, 2004).
\bibitem{ot} {\em Semiconductors---Basic Data}, edited by Otfried Madelung (Springer, Berlin, 1996), p.p. 151;
E. T. Yu, J. O. McCaldin, and T. C. McGill, Solid State Phys. {\bf 46}, 
1 (1992).
\bibitem{lei} X. L. Lei, J. L. Birman, and C. S. Ting,
J. Appl. Phys. {\bf 58}, 2270 (1985).
\bibitem{rhall}M.  Lundustrom,
  {\em Fundamentals of the Carrier Transport}
(Cambridge University Press, Cambridge, England, 2000);
J. Singh, {\em Physics of the Semiconductors and Their
  Heterostructures}  (McGraw-Hill, New York, 1993).
\bibitem{aronov} A. G. Aronov, G. E. Pikus, and A. N. Titkov,
Zh. \'Eksp. Teor. Fiz. {\bf 84}, 1170 (1983) [Sov. Phys. JETP {\bf 57}, 680
(1983)].
\bibitem{marush}V. A. Marushchak, M. N. Stepanova, and A. N.
Tikov, Fiz. Tverd. (Donetsk) {\bf 25}, 3537 (1983).
\bibitem{juss} B. Jusserand, D. Richards, G. Allen, C. Priester,
and B. Etienne, Phys. Rev. B {\bf 51},
  4707 (1994).
\bibitem{gore}A. T. Gorelenko, B. A. Marushchak, and A. N. Tikov.
Izv. Akad. Nauk SSSR, Ser. Fiz. {\bf
    50}, 290 (1986).
\bibitem{cardona} M. Cardona, N. E. Christensen, and G. Fasol,
Phys. Rev. B {\bf 38}, 1806 (1987).
\bibitem{chan} A. N. Chantis, M. van Schilfgaarde, and T. Kotani, Phys. Rev. Lett. {\bf 96},
  086405 (2006).
\bibitem{zhushi1}It is difficult to give an exact analytical
expression of $\tau_p^{ee}$  as we take into account of the
Coulomb scattering to all orders as well as the counter effect of
the Coulomb scattering to the inhomogeneous broadening. However, one can
roughly estimate $\tau_p^{ee}$ by considering the Coulomb
scattering to the lowest order (2nd order) and further
ignoring the counter effect in the degenerate and
non-degenerate limits following the approach in
Ref.\ \onlinecite{glazov}.
\bibitem{yu} P. Y. Yu and M. Cardona, {\em Fundamentals of Semiconductors}
(Springer, Berlin, 2003), 3rd Ed., p.\ 222.
\bibitem{glazov} M. M. Glazov and E. L. Ivchenko, Zh. \'{E}ksp.
Teor. Fiz. {\bf 126} 1465 (2004) [JETP {\bf 99}, 1279 (2004)].
\bibitem{glazov1} M. M. Glazov and E. L. Ivchenko, Pis'ma Zh.
\'{E}ksp. Teor. Fiz. {\bf 75}, 476 (2002) [JETP Lett. {\bf 75}, 403 (2002)].
\bibitem{comment} In the weak scattering regime, the Coulomb scattering can
also reduce the spin R/D time.\cite{lue}
\bibitem{harley3} Private communication with R. T. Harley.
\bibitem{comment1} It is stressed here that
unlike the high temperature case where the preparation of the
steady-state initial electron distributions under strong electric field
proposed in Ref.\ \onlinecite{wu8} is not essential as the time scale
to reach the steady state after turning on the electric field is
negligible in comparison with the time scale of the spin R/D,
at low temperature case due to the weak electron-AC phonon scattering,
the preparation is essential as  both time scales can be
comparable to each other.
\bibitem{run} A. P. Dmitriev, V. Yu Kachorovskii, M. S. Shur, and M. Stroscio,
Solid State Commun. {\bf 113}, 565 (2000); A. P. Dmitriev, V. Yu Kachorovskii,
and M. S. Shur, J. Appl. Phys. {\bf 89}, 3793 (2001).
\bibitem{conwell} E. M. Conwell, {\em High Field Transport in Semiconductors}
(Academic press, New York and London, 1967), p.\ 21.
\end{thebibliography}
\end{document}